\documentclass[conference]{IEEEtran}
\IEEEoverridecommandlockouts

\usepackage{cite}
\usepackage{graphicx}
\usepackage{times}
\usepackage{latexsym}

\usepackage[mathscr]{eucal}
\usepackage{array}
\usepackage{fancyhdr}
\usepackage{amsmath,amssymb}
\usepackage{bm} 
\usepackage{subfigure}
\usepackage{citesort}
\usepackage{multirow}

\ifCLASSINFOpdf
\else
\fi

\newcommand{\mat}[1]{{\uppercase{\mathbf{#1}}}}

\newcommand{\diag}{{\rm{diag}}}
\renewcommand{\H}{\mat{H}} % ''accent
\newcommand{\I}{\mat{I}}

\newcommand{\X}{\mat{X}}
\newcommand{\Y}{\mat{Y}}

\usepackage{xcolor}

\makeatother

\usepackage{cite}
\usepackage{amsmath,amssymb,amsfonts}
\usepackage{algorithmic}
\usepackage{graphicx}
\usepackage{textcomp}
\usepackage{xcolor}

\def\BibTeX{{\rm B\kern-.05em{\sc i\kern-.025em b}\kern-.08em
    T\kern-.1667em\lower.7ex\hbox{E}\kern-.125emX}}

\begin{document}

\title{On the RIS Manipulating Attack and Its Countermeasures in Physical-layer Key Generation
}

%% author ----
\author{
	\IEEEauthorblockN{Lei Hu$^{*}$, Guyue Li$^{*\dag}$, Hongyi Luo$^{*}$, Aiqun Hu$^{\dag\ddag}$}
	\IEEEauthorblockA{ $^{*}$School of Cyber Science and Engineering, Southeast University, Nanjing, 210096, China}
	\IEEEauthorblockA{ $^{\dag}$Purple Mountain Laboratories for Network and Communication Security, Nanjing, 210096, China}
	\IEEEauthorblockA{ $^{\ddag}$National Mobile Communications Research Laboratory, Southeast University, Nanjing, 210096, China}
%	\IEEEauthorblockA{\{lei-hu, guyuelee,aqhu@seu.edu.cn\}@seu.edu.cn, hongyiluo\_seu@163.com}
	\IEEEauthorblockA{Corresponding author: Guyue Li, Email: {guyuelee}@seu.edu.cn}
}

\maketitle

\begin{abstract}
	Reconfigurable Intelligent Surface (RIS) is a new paradigm that enables the reconfiguration of the wireless environment. Based on this feature, RIS can be employed to facilitate Physical-layer Key Generation (PKG). However, this technique could also be exploited by the attacker to destroy the key generation process via manipulating the channel features at the legitimate user side. Specifically, this paper proposes a new RIS-assisted Manipulating attack (RISM) that reduces the wireless channel reciprocity by rapidly changing the RIS reflection coefficient in the uplink and downlink channel probing step in orthogonal frequency division multiplexing (OFDM) systems. The vulnerability of traditional key generation technology based on channel frequency response (CFR) under this attack is analyzed. Then, we propose a slewing rate detection method based on path separation. The attacked path is removed from the time domain and a flexible quantization method is employed to maximize the Key Generation Rate (KGR). The simulation results show that under RISM attack, when the ratio of the attack path variance to the total path variance is 0.17, the Bit Disagreement Rate (BDR) of the CFR-based method is greater than 0.25, and the KGR is close to zero. In addition, the proposed detection method can successfully detect the attacked path for SNR above 0 dB in the case of 16 rounds of probing and the KGR is 35 bits/channel use at 23.04MHz bandwidth.
\end{abstract}

\begin{IEEEkeywords}
	Physical layer security, Reconfigurable Intelligent Surface, OFDM, secret key generation, active attack.
\end{IEEEkeywords}
\IEEEpeerreviewmaketitle

\section{Introduction}
\label{section1}
The rapid development of the fifth-generation (5G) communication system has greatly increased the amount of data transmitted in the air interface. However, due to the broadcast characteristics of the wireless media, a large amount of confidential information may be eavesdropped on by unauthorized users. %\cite{7414384}. 
The Physical-layer Key Generation (PKG) uses the reciprocity of the wireless channel to generate a pair of symmetric secret key to encrypt data. Due to the time-varying and spatial uncorrelation of the channel, the eavesdropper cannot predict the channel of the legitimate user and secure one-time pad communication can be realized \cite{18Tcom}. 

% introduction to RIS
However, the performance of PKG is limited by channel reciprocity and time-varying. Recently, Reconfigurable Intelligent Surface (RIS) emerges as a new technology that can realize the regulation of electromagnetic waves and thus change the wireless propagation environment \cite{Wu-Beamforming}. Due to the passive characteristics and low hardware cost of RIS, RIS-assisted wireless communication system has been studied extensively. 

Based on the above features, little existing literature has studied RIS-assisted PKG. Ji \textit{et al.} \cite{JiOTP} randomly change the phase of RIS to introduce artificial randomness and increase the key generation rate (KGR). In their other work \cite{Ji_IRS}, RIS reflecting coefficients are optimized to maximize the lower bound of secret key rate in multiple eavesdroppers scenario. 
Moreover, Lu \textit{et al.} \cite{SPletter} adjusted the placement of RIS to maximize the key rate capacity. 
The above literature is all based on the single antenna scenario. Recently, \cite{chen2021intelligent} studies optimization of RIS beamforming to maximize key rate in MISO system.
However, these works are based on the assumption that RIS is controlled by the legitimate party, e.g. BS, ignoring the fact that it could also be controlled by the attacker.

As far as we know, it is the first time to investigate the RIS in secret key generation from the aspect of active attack. In this paper, we propose a new RIS-assisted Manipulating attack (RISM) in the orthogonal frequency division multiplexing (OFDM) system and give the RISM detection method and key generation method based on path separation. The main contributions of this paper are summarized as follows. 
\begin{itemize}
	\item An RISM method is proposed to reduce the KGR, in which the active attacker Eve rapidly changes the phase of RIS to manipulate the wireless environment. The vulnerability of traditional key generation technology using the channel frequency response (CFR) coefficient under this attack is analyzed.
	
	\item A slewing rate detection method based on path separation and multiple channel probing is proposed. Then, we adopt a flexible quantization method based on the separated paths to improve the KGR.
	
	\item Simulation results show that the KGR of CFR coefficient under RISM is close to zero when the ratio of the variance of the attacked path is 0.17. The success rate of detection is close to 1 for SNR above 0 dB. Also, the KGR is improved significantly.
	
\end{itemize}

\section{The OFDM-based PKG System}

We first consider a general key generation model in OFDM system.
It is assumed that Alice and Bob are equipped with single antenna. To leverage the channel reciprocity of time-division duplex (TDD) mode, Alice and Bob take turns to estimate channels within coherence time.
During the channel probing stage, the OFDM symbol transmitted by Alice is denoted as  $\X=\diag(x_1,x_2,\cdots,x_L)$, where $L$ is the number of subcarriers.
The received signal at Bob side after removing cyclic prefix is represented as 
\begin{align}
\Y_b = \X \H_{d} + \mathbf{N}_b
\end{align}
where $\H_{d}\in \mathbb{C}^{L \times 1}$ denotes the direct channel from Alice to Bob, 
$\mathbf{N}_b\sim \mathcal{N}_{c}\left(\mathbf{0}, \sigma^{2}_n \mathbf{I}_{L}\right)$ is the complex additive white Gaussian noise.
Then, the least-square (LS) estimation is performed at Bob and the estimated channel is given by
\begin{align}
\bar{\mathbf{H}}_b=\mathbf{H}_d+  \tilde{\mathbf{N}}_b
\end{align}
where the noise component $\tilde{\mathbf{N}}_b=\X^{-1}\mathbf{N}_b$ and 
 $\tilde{\mathbf{N}}_b\sim \mathcal{N}_{c}\left(\mathbf{0}, \sigma^{2}_n\X^{-1} \right)$.
For the purpose of exposition, we assume $x_1=x_2=\cdots=x_L=1$. 
Since uplink and downlink channel probing is completed within coherence time, the channel reciprocity holds between channel probing. 
Similarly, Alice estimates the channel as
\begin{align}
\bar{\mathbf{H}}_a=\mathbf{H}_d+ \tilde{\mathbf{N}}_a
\end{align}
Then, in the quantization step, Cumulative Distribution Function (CDF) based quantization method can be used to convert channel coefficients into bits \cite{2018trans}. Due to the existence of noise, there will be a small amount of inconsistents bits between Alice's and Bob's quantization results, which can be corrected through information reconciliation step. Finally, the information leaked in the information reconciliation step can be eliminated through the privacy amplification.

\section{RISM Attack Scheme}
In this section, we propose a novel RIS-assisted attack method, RISM, which reduces the channel reciprocity by changing the reflection coefficient of the uplink and downlink channel probing and thus reduces the KGR. We then give the change of channel correlation coefficient and the influence on CDF based quantization after RISM.
\subsection{RISM Attack Model}
\begin{figure}
	\centering
	\includegraphics[width=3.2in]{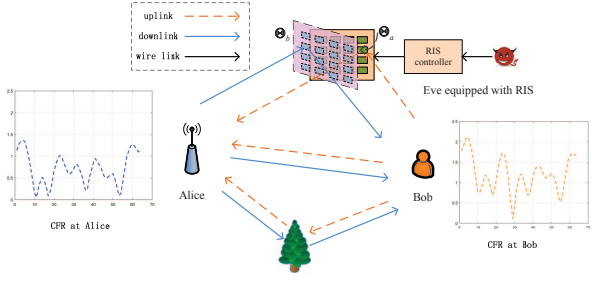}
	\caption{The proposed RISM attack model.}
	\label{system model}
\end{figure}

Fig. \ref{system model} shows an RISM attack in PKG system, wherein a malicious attacker Eve employs an RIS to destroy the reciprocal channel between Alice and Bob. The RIS controlled by Eve has $M$ passive reflection elements, and the phase of each element in the uplink channel probing stage is different from that of the downlink stage.
We consider an active attack scheme where Eve doesn't require extra hardware to eavesdrop and doesn't need to know the time slot of the channel probing. Particularly, Eve randomly and rapidly changes the reflection coefficient of each element in both uplink and downlink. The time interval of change is much smaller than the coherence time. Thus, the channel estimated by Bob in RISM is
\begin{align}
\widehat{\H}_b &= \mathbf{H}_d+\sum_{m=1}^{M} \theta_{b,m} \mathbf{H}_{ar,m} \odot \mathbf{H}_{rb,m} + \tilde{\mathbf{N}}_b \\
&=\mathbf{H}_d+  \mathbf{H}_{r}\mathbf{\Theta}_{b}  + \tilde{\mathbf{N}}_b
\end{align}
where $\odot$ denotes the Hadamard product,  $\mathbf{H}_{ar,m} \in \mathbb{C}^{L \times 1}$ and $\mathbf{H}_{rb,m} \in \mathbb{C}^{L \times 1}$ are the links from Alice to $m$-th sub-surface of RIS and from the $m$-th sub-surface to Bob, respectively. 
$\mathbf{H}_{r}\mathbf{\Theta}_{b}$ is the equivalent cascaded channel matrix and $\mathbf{H}_{r}=[\mathbf{H}_{r,1} ,\mathbf{H}_{r,2}, \cdots,\mathbf{H}_{r,M}]$, where  $\mathbf{H}_{r,m}\triangleq\mathbf{H}_{ar,m} \odot \mathbf{H}_{rb,m}, m=1,2,\cdots, M $. The RIS reflection vector is  $\mathbf{\Theta}_{b}=[\theta_{b,1},\theta_{b,2},\cdots,\theta_{b,M}]^T$.
Similarly, the channel eatimation at Alice in RISM can be expressed as
\begin{align}
	\widehat{\H}_a&=\mathbf{H}_d+  \mathbf{H}_{r}\mathbf{\Theta}_{a}  + \tilde{\mathbf{N}}_a
\end{align}
where the reflection coefficient $\theta_{u,m}= e^{j \phi_{u,m}}, u=a,b $ and the phase $\phi_{u,m}$ follows a independent and identically distributed (i.i.d.) uniform distribution over $[0,2 \pi)$.

\subsection{Evaluation of RISM Effect}
To evaluate the impact of RISM on channel reciprocity, Pearson’s cross-correlation coefficient of CFR can be calculated. 
The $\ell$-th element of $\widehat{\H}_{u}$ is expressed as $\widehat{H}_{u,\ell}=H_{d,\ell} + \sum_{m=1}^{M}\theta_{u,m}H_{r,m}^{\ell}+\tilde{N}_{u,\ell}$. 
Hence, the correlation coefficient of $\ell$-th channel estimation in the frequency domain is
\begin{align}
\label{before}
&\rho_{\ell} =\frac{\mathbb{E}\{(\widehat{H}_{a,\ell}-\mu_{a,\ell})(\widehat{H}_{b,\ell}-\mu_{b,\ell})^*\}}{\sqrt{\mathbb{E}\{|\widehat{H}_{a,\ell}-\mu_{a,\ell}|^2\}\mathbb{E}\{|\widehat{H}_{b,\ell}-\mu_{b,\ell}|^2\}}}\\
&=\frac{\sigma_{d,\ell}^{2} + G_1}{\sigma_{d,\ell}^{2}+G_2+\xi_\ell+\sigma_{n,\ell}^2} \\
&=\frac{\sigma_{d,\ell}^{2}}{\sigma_{d,\ell}^{2} + \xi_\ell + \sigma_{n,\ell}^{2}}  
\end{align}
where $\sigma_{d,\ell}^{2}$ and $\sigma_{n,\ell}^2$ denote the variance of direct channel and noise in the $\ell$-th subcarrier, respectively. Since the reflection phase is uniformly distributed, we have 
\begin{align}
	G_1&=\mathbb{E}\{(\widehat{H}_{d,\ell}-\mu_{d,\ell})(\sum_{m=1}^{M}\theta_{b,m}H_{r,m}^{\ell})^{*}   \} \\ \notag
	&+\mathbb{E}\{(\widehat{H}_{d,\ell}-\mu_{d,\ell})^{*}(\sum_{m=1}^{M}\theta_{a,m}H_{r,m}^{\ell})   \} \\ \notag
	&+\mathbb{E}\{(\sum_{m=1}^{M}\theta_{b,m}H_{r,m}^{\ell})^{*}(\sum_{m=1}^{M}\theta_{a,m}H_{r,m}^{\ell})   \} =0 \notag\\ 
G_2&=\mathbb{E}\{2\Re((\widehat{H}_{d,\ell}-\mu_{d,\ell})(\sum_{m=1}^{M}\theta_{a,m}H_{r,m}^{\ell})^{*})   \} =0
\end{align} 
Also, $\xi_\ell = \sum_{m=1}^{M}\sigma_{r,m,\ell}^{2}$ is the sum of the cascaded channel variances. 
Since the variance $\xi_{\ell} \textgreater 0$, RISM will inevitably lead to a decrease in the CFR correlation coefficient. 
As the $\xi_{\ell}$ increases, the correlation coefficient tends to zero, resulting in low channel reciprocity.

In addition, to reflect the impact of RISM on quantization step in PKG, we analyze the CDF based bit disagreement rate (BDR) of CFR under this attack,  
where Alice and Bob use the CDF of channel coefficient to divide its range into $2^{m_{i}}$ equally spaced regions and gray code is used to encode $\widehat{H}_{u,\ell}$ into $m$ bit(s).
The BDR after RISM can be approximated as formula (\ref{BDR_attack}) in the case of low bit disagreement, 
\begin{figure*}
	\begin{align}
	\label{BDR_attack}
	P_{B D}^{'} &\approx \frac{1}{m_i}\{1-P_{C A}\}\\ \notag
	&=  \frac{1}{m_i}\left\{1-\int_{v=-\infty}^{\infty}\left\{\Phi\left[\frac{(\sigma_{d,\ell}^{2}+ A)\Phi^{-1}\left[\alpha_{a}\right]-\sigma_{d,\ell}^{2} v}{\sqrt{(2\sigma_{d,\ell}^{2}+A) A}}\right]
	-\Phi\left[\frac{(\sigma_{d,\ell}^{2}+ A)\Phi^{-1}\left[\beta_{a}\right]-\sigma_{d,\ell}^{2} v}{\sqrt{(2\sigma_{d,\ell}^{2}+A) A}}\right]\right\} \frac{e^{-v^{2} / 2}}{\sqrt{2 \pi}} d v\right\}
	\end{align}
\end{figure*}
where $A=\sigma_{d,\ell}^{2}+\xi_\ell + \sigma_{n,\ell}^{2}$, $\Phi$ is the CDF of a Gaussian distribution with zero mean unit variance, $\alpha_{a}$ and $\beta_{a}$ are the upper and lower bounds of $\widehat{H}_{a,\ell}$ in the agreement area when $\widehat{H}_{a,\ell}$ is given, i.e.
\begin{align}
\alpha_{a}&=\text{min}\{1,\lceil F(\widehat{H}_{a,\ell})2^{-m_i} \rceil 2^{-m_i}\}    \\
\beta_{a}&=\text{max}\{0,\lfloor F(\widehat{H}_{a,\ell})2^{-m_i} \rfloor 2^{-m_i}\} 
\end{align} 

According to the results of literature \cite{2018trans} and formula (\ref{BDR_attack}), the BDR will increase with the increase of variance of the cascaded channel. 
Considering that the error-correcting ability of information reconciliation is limited, 
when the BDR is beyond the reconciliation capability, the users have to discard the generated bits, indicating the failure of this round of secret key generation.

\section{RISM detection and Countermeasures}
In this section, we propose a slewing rate detection method and a flexible quantization method to resist RISM attack. In addition, we give the theoretical key rate performance loss when the OFDM bandwidth is insufficient to distinguish each path.

\subsection{Path Separation and Detection Method}
We consider that in a multi-path environment, only the paths through the RIS change rapidly, while others remain the same during the coherence time.
Motivated by this, by separating the attacked paths, the impact of the attack can be eliminated and we can reconstruct the reciprocal channel for key generation.
In OFDM system, each path can be obtained via an $L$-point inverse discrete Fourier transform (IDFT) \cite{18CL}.
Given the proximity of adjacent elements of the RIS, it is often difficult to distinguish multiple paths through different elements. Therefore, it can be assumed that only one path passes through the RIS \footnote{Assuming a bandwidth of $B$ = 20 MHz in IEEE 802.11a, paths can only be distinguished if the distance between them is greater than $d=c/B=15m$. However, most RIS sizes are much smaller than $d$.}.
The matrix form of the path separation can be described as
\begin{align}
\mathbf{h}_u&=\frac{1}{L}\mathbf{F}_L^H(\mathbf{H}_d+  \mathbf{H}_{r}\mathbf{\Theta}_{u}  + \tilde{\mathbf{n}}_u) \\ 
& =\mathbf{h}_d + \mathbf{h}_{r,u} + \bar{\mathbf{n}}_u
\end{align}
where $\mathbf{F}_L$ refers	to the $L$ order discrete Fourier transform (DFT) matrice.
Since the number of paths is small in the multipath channel, it is generally considered sparse. Therefore, the direct channel and the cascade channel can be characterized as $\mathbf{h}_d=\left[\bar{h}_{1}, \ldots, \bar{h}_{L_d}, \mathbf{0}_{1 \times(L-L_d)}\right]^{T}$ and 
$\mathbf{h}_{r,u}=\left[0, \ldots, 0,\bar{h}_{u,k}, \mathbf{0}_{1 \times(L-k)}\right]^{T}$, respectively.  The $k$-th element of $\mathbf{h}_{r,u}$ is 
the attacked path.

Then, we distinguish which paths are available for key generation and which path is attacked by Eve.
We propose to detect RISM by sounding the channel multiple times in a coherence time, which can also reduce noise on unattacked paths. 
Specifically, according to the time chart of proposed key generation based on IEEE 802.11a Data frame, a round of channel probing can be carried out every 2.8ms, providing the possibility of attack detection. 
In order to minimize the impact of noise, the detection method is shown in (\ref{detection})  
\begin{align}
	\label{detection}
	\frac{\sum_{q=1}^{Q} h_{i,q}h_{i,q}^{*}}{Q}-\frac{1}{L_d+1}\sum_{\ell=1}^{L_d+1}\frac{\sum_{q=1}^{Q} h_{\ell,q}h_{\ell,q}^{*}}{Q}
	 \underset{\mathcal{H}_{1}}{\overset{\mathcal{H}_{0} }{\gtrless}} \alpha
\end{align}
 where $h_{i,q}$ is the $q$-th round channel probing on $i$-th path and $Q$ is the number of probing. 
 $\mathcal{H}_{0}$ indicates the $i$-th path is attacked by RISM and $\mathcal{H}_{1}$ indicates the RISM is absent  on $i$-th path. 
Note that $\alpha$ depends on the variance of the attacked path and is set to be an empirical value.
With the increase of the number of probes, the estimation of the variance will be more accurate. However, the channel probing will result in the increase of pilot overhead and the number of channel probing rounds is limited in the coherence time. Therefore, a tradeoff should be made between the number of channel probing and the success rate.
Finally, Alice and Bob will discard the attacked path, and the remaining paths are perfectly satisfied channel reciprocity and can be used for key generation.  

\subsection{Path Separation based Flexible Quantization}
Considering that after IDFT is performed, different paths will have different signal-to-noise ratios (SNR). To make full use of high SNR paths and reduce the BDR of low SNR paths, we use a flexible quantization method to maximize the generated key bits. 
The SNR in the frequency domain is defined as 
\begin{align}
\text{SNR} &= \frac{\mathbb{E}\{(\mathbf{H}_d+  \mathbf{H}_{r}\mathbf{\Theta}_{u})^H(\mathbf{H}_d+  \mathbf{H}_{r}\mathbf{\Theta}_{u})\}}{\mathbb{E}\{\tilde{\mathbf{N}}_u^H\tilde{\mathbf{N}}_u\}}\\
&=\frac{\mathbb{E}\{\mathbf{H}_d^{H}\mathbf{H}_d\}+  M \mathbb{E}\{\mathbf{H}_{r}^H \mathbf{H}_{r}\}}{L\sigma_{n}^2}
\end{align}
After the RISM attack detection, the path through RIS is removed. Then, the SNR of the $i $-th path is \cite{18Tcom}
\begin{align}
\text{SNR}_{i} &= \frac{\mathbf{f}_{i}/L\mathbb{E}\{\mathbf{H}_d\mathbf{H}_d^H\}\mathbf{f}_{i}^H/L}{\sigma_{n}^2/L}\\
&=\frac{\mathbf{f}_{i}/L\mathbf{F}_L\mathbb{E}\{\mathbf{h}_d\mathbf{h}_d^H\}\mathbf{F}_L^H \mathbf{f}_{i}^H/L}{\sigma_{n}^2/L}\\
&=\frac{L\mathbb{E}\{|\bar{h}_{i}|^2\}}{\sigma_{n}^2}, \quad i\neq k
\end{align}
where $\mathbf{f}_{i}$ is the $i$-th row of the DFT matrix $\mathbf{F}$.
In the practical system, in order to ensure the key consistency, the key error rate (KER) needs to be below $10^{-3}$. To meet this requirement, 
we set quantization levels according to the SNR of each path. Particularly, each path will use higher quantization bits to increase the key rate when it is lower than a given KER.

\subsection{Performance Analysis of Key Generation}
Due to the finite bandwidth, the resolution
to estimate the delay of every path is finite as well.
Particularly, under bandwidth $B$, the path delay resolution is expressed as $\Delta \tau=\frac{1}{B}$. As a result, 
when the bandwidth is insufficient to resolve each path, paths that are in the same tap as the attack path will be discarded after the attack detection, which affects the KGR.

In order to measure the impact of the discarded available paths on the KGR, we assume that there are $N$ discarded paths in the total path $L_d$, and $N$ increases with the reduction of bandwidth. 
We assume that 
$\tilde{\mathbf{h}}_{u}\in \mathbb{C}^{N\times 1}$ denotes the discarded paths. 
At this point, the theoretically reduced key rate is
\begin{align}
R_{reduced} &= I(\tilde{\mathbf{h}}_{a};\tilde{\mathbf{h}}_{b}) \\
& = \text{log}_{2} \frac{|\bm{R}_{a}||\bm{R}_{b}|}{|\bm{R}_{a}||\bm{R}_{b}-\bm{R}_{ab}\bm{R}_{a}^{-1}\bm{R}_{ba}|} \\
& = \text{log}_{2} \frac{|\bm{\Lambda}_{h}+\frac{\sigma_{n}^2}{L}\I|}{|\bm{\Lambda}_{h}+\frac{\sigma_{n}^2}{L}\I-\bm{\Lambda}_{h}(\bm{\Lambda}_{h}+\frac{\sigma_{n}^2}{L}\I)^{-1}\bm{\Lambda}_{h}^H|}\\
&=\sum_{i=1}^{N}\text{log}_{2}\frac{1}{1-(\frac{L\lambda_{i}/\sigma_{n}^2}{1+L\lambda_{i}/\sigma_{n}^2})^2}
\end{align}
where the covariance matrix
\begin{align}
\bm{R}_{b}&=\mathbb{E}\{\tilde{\mathbf{h}}_{b}\tilde{\mathbf{h}}_{b}^H\} =\bm{\Lambda}_{h}+\sigma_{n}^2\I/L \\
\bm{R}_{ab}&=\mathbb{E}\{\tilde{\mathbf{h}}_{a}\tilde{\mathbf{h}}_{b}^H\} =\bm{\Lambda}_{h}
\end{align} 
where $\lambda_{i}=\left[\bm{\Lambda}_{h}\right]_{i,i}$ is the variance of $i$-th discarded path. 
We notice that the reduced key rate depends on the path variance and the number of paths discarded. Hence, higher bandwidth is needed to improve KGR.

\section{Numerical Results}
% 仿真环境参数
\begin{figure}
	\centering
	\includegraphics[width=3.2in]{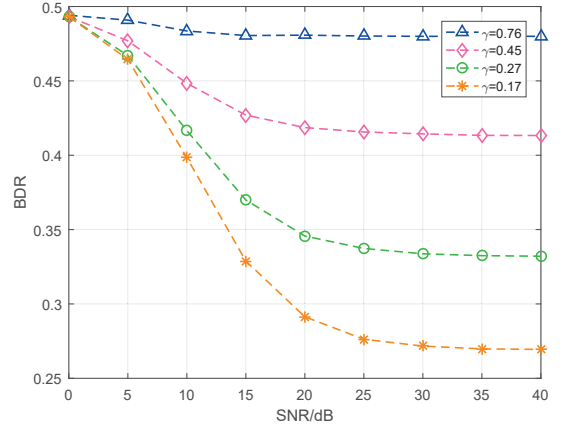}
	\caption{BDR of CFR amplitude quantization after RISM.}
	\label{BDR_SNR_1}
\end{figure}
In this section, we evaluate the performance of our proposed
RISM attack and countermeasures via numerical results. The geometry-based
channel model 3GPP Spatial Channel Model (SCM) is adopted for simulation. In the simulation of the broadband multipath environment, the number of paths before being attacked is set to 7, and each path has 20 subpaths. 
The number of RIS reflection elements $M$ is 30, and the distance between adjacent elements is $\lambda/2$. The positions of Alice, Bob, and RIS are randomly distributed, and the distance between them is evenly distributed within [35, 200] m.
In OFDM system, the Carrier frequency $f_c$ = 2 GHz and 64 subcarriers are implemented. The bandwidth $B$ is set to be 23.04 MHz and each path can be resolved.

Considering that as the number of RIS reflection elements increases, the subpath of the attacked path increases, increasing the variance of the attacked path. 
Define $\gamma=\frac{\mathbb{E}\{|\bar{h}_{u,k}|^2\}}{\sum_{i=1}^{L_d}\mathbb{E}\{|\bar{h}_{i}|^2\}+\mathbb{E}\{|\bar{h}_{u,k}|^2\}}$ as the ratio of the variance of the attacked path to the variance of the total paths.
Fig. \ref{BDR_SNR_1} compares the BDR of the CFR v.s. different $\gamma$ after the RISM, where the quantization bit $m=1$. We can find that as the SNR increases, the BDR corresponding to different $\gamma$ gradually decreases and tends to a constant value. 
However, we note that the BDR increases with the increase of $\gamma$. 
When $\gamma$ is small and equal to 0.17, the final BDR tends to 0.27. According to the literature \cite{5205475}, the maximum tolerable BDR is 0.11 and the generated bits in RISM will be discarded.
\begin{figure}
	\centering
	\includegraphics[width=3.2in]{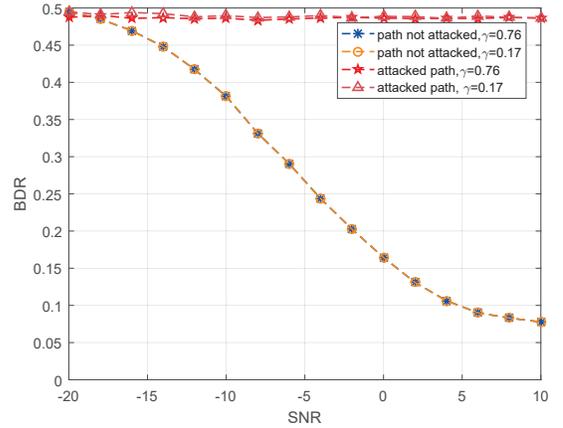}
	\caption{BDR of different path versus SNR.}
	\label{BDR_SNR_2}
\end{figure}

Fig. \ref{BDR_SNR_2} shows the BDR of different paths after the attack, where the BDR of the attacked path is around 0.5 at different $\gamma$, which means that the path is completely non-reciprocal. Besides, the path that does not go through RIS decreases with the increase of SNR, 
providing conditions for the reconstruction of the reciprocal channel.
\begin{figure}
	\centering
	\includegraphics[width=3.2in]{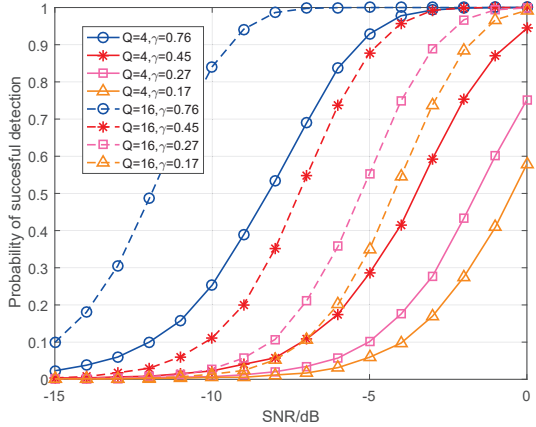}
	\caption{Probability of successful detection versus SNR and probing times.}
	\label{Detect_success1}
\end{figure}

Fig. \ref{Detect_success1} shows the probability of successful detection. 
As the variance of noise decreases, the probability increases. In addition, the success rate increases with the increase of attacked path variance ratio $\gamma$. Also, as the number of probes increases, a higher detection probability can be obtained. In the simulation setup, assume the speed $v=1m/s$, then the Doppler spread $D_s$ is about 6.7 Hz. Thus, the coherence time is $T_c=\frac{1}{2D_s}=75$ms. Since multiple probing rounds increase the pilot overhead, we set the round of probing $Q=16$ to achieve a tradeoff between the success rate and the overhead.

\begin{figure}
	\centering
	\includegraphics[width=3.2in]{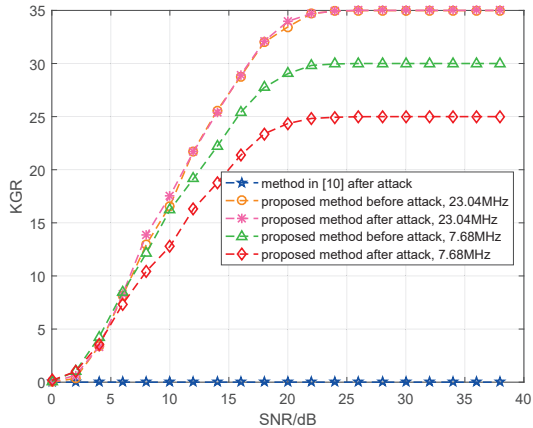}
	\caption{KGR comparison between the CFR method and the proposed method.}
	\label{SKG_bandwidth}
\end{figure}
Fig. \ref{SKG_bandwidth} presents the practical key rate after channel probing, CDF-based quantization, and  low-density parity-check code (LDPC) bit-flipping-based information reconciliation. 
We notice that the BDR of traditional key generation method \cite{OFDM_2008} based on CFR is high, resulting in low KGR.
With the increase of SNR, the number of quantization bits of each path increases, leading to the increase of total key bits generated.
At the same time, at 7.68MHz bandwidth, one available path and the attacked path are in the same tap. Therefore, a reciprocal path is discarded and the KGR after the attack is lower than that before the attack. 
In contrast, with high bandwidth of 23.04MHz, the path resolution is high enough that the attacked path can be removed independently. Therefore, the nearly perfect key rate can be achieved.

\section{Conclusion}
In this paper, an RISM scheme aiming to reduce the rate of key generation was proposed and analyzed. 
In traditional CFR-based method, we showed its vulnerability on the channel reciprocity and quantization step of key generation under RISM. 
Therefore, the CFR-based method cannot generate effective key bits due to its high BDR. For more, the slewing rate detection method and flexible quantization were proposed to remove the attacked path and improve the KGR. Numerical results showed that a high success rate of detection can be obtained for SNR above 0 dB and a high KGR can be achieved.

\appendices

\section*{Acknowledgment}
This work was supported in part by the National Natural Science Foundation of China
under Grant 61941115 and Grant 61801115, in part by the Jiangsu key R \& D plan BE2019109, and in part by the Zhishan Youth Scholar Program of SEU (3209012002A3).

\bibliographystyle{IEEEtran}
\bibliography{IEEEabrv,mmm}

\begin{thebibliography}{10}
\providecommand{\url}[1]{#1}
\csname url@samestyle\endcsname
\providecommand{\newblock}{\relax}
\providecommand{\bibinfo}[2]{#2}
\providecommand{\BIBentrySTDinterwordspacing}{\spaceskip=0pt\relax}
\providecommand{\BIBentryALTinterwordstretchfactor}{4}
\providecommand{\BIBentryALTinterwordspacing}{\spaceskip=\fontdimen2\font plus
\BIBentryALTinterwordstretchfactor\fontdimen3\font minus
  \fontdimen4\font\relax}
\providecommand{\BIBforeignlanguage}[2]{{%
\expandafter\ifx\csname l@#1\endcsname\relax
\typeout{** WARNING: IEEEtran.bst: No hyphenation pattern has been}%
\typeout{** loaded for the language `#1'. Using the pattern for}%
\typeout{** the default language instead.}%
\else
\language=\csname l@#1\endcsname
\fi
#2}}
\providecommand{\BIBdecl}{\relax}
\BIBdecl

\bibitem{18Tcom}
G.~Li, A.~Hu, J.~Zhang, L.~Peng, C.~Sun, and D.~Cao, ``High-agreement
  uncorrelated secret key generation based on principal component analysis
  preprocessing,'' \emph{IEEE Transactions on Communications}, vol.~66, no.~7,
  pp. 3022--3034, 2018.

\bibitem{Wu-Beamforming}
Q.~{Wu} and R.~{Zhang}, ``Intelligent reflecting surface enhanced wireless
  network via joint active and passive beamforming,'' \emph{IEEE Transactions
  on Wireless Communications}, vol.~18, no.~11, pp. 5394--5409, 2019.

\bibitem{JiOTP}
Z.~{Ji}, P.~L. {Yeoh}, G.~{Chen}, C.~{Pan}, Y.~{Zhang}, Z.~{He}, H.~{Yin}, and
  Y.~{Li}, ``Random shifting intelligent reflecting surface for otp encrypted
  data transmission,'' \emph{IEEE Wireless Communications Letters}, pp. 1--1,
  2021.

\bibitem{Ji_IRS}
Z.~{Ji}, P.~L. {Yeoh}, D.~{Zhang}, G.~{Chen}, Y.~{Zhang}, Z.~{He}, H.~{Yin},
  and Y.~{li}, ``Secret key generation for intelligent reflecting surface
  assisted wireless communication networks,'' \emph{IEEE Transactions on
  Vehicular Technology}, vol.~70, no.~1, pp. 1030--1034, 2021.

\bibitem{SPletter}
X.~{Lu}, J.~{Lei}, Y.~{Shi}, and W.~{Li}, ``Intelligent reflecting surface
  assisted secret key generation,'' \emph{IEEE Signal Processing Letters}, pp.
  1--1, 2021.

\bibitem{chen2021intelligent}
Y.~{Chen}, G.~{Li}, C.~{Pan}, L.~{Hu}, and A.~{Hu}, ``{Intelligent Reflecting
  Surface-Assisted Secret Key Generation In Multi-antenna Network},''
  \emph{arXiv e-prints}, p. arXiv:2105.00511, May 2021.

\bibitem{2018trans}
N.~{Patwari}, J.~{Croft}, S.~{Jana}, and S.~K. {Kasera}, ``High-rate
  uncorrelated bit extraction for shared secret key generation from channel
  measurements,'' \emph{IEEE Transactions on Mobile Computing}, vol.~9, no.~1,
  pp. 17--30, 2010.

\bibitem{18CL}
G.~Li, A.~Hu, C.~Sun, and J.~Zhang, ``Constructing reciprocal channel
  coefficients for secret key generation in fdd systems,'' \emph{IEEE
  Communications Letters}, vol.~22, no.~12, pp. 2487--2490, 2018.

\bibitem{5205475}
D.~Elkouss, A.~Leverrier, R.~Alleaume, and J.~J. Boutros, ``Efficient
  reconciliation protocol for discrete-variable quantum key distribution,'' in
  \emph{2009 IEEE International Symposium on Information Theory}, 2009, pp.
  1879--1883.

\bibitem{OFDM_2008}
S.~{Yasukawa}, H.~{Iwai}, and H.~{Sasaoka}, ``Adaptive key generation in secret
  key agreement scheme based on the channel characteristics in ofdm,'' in
  \emph{2008 International Symposium on Information Theory and Its
  Applications}, 2008, pp. 1--6.

\end{thebibliography}


\end{document}